\documentclass[12pt]{article}

\usepackage{amssymb,amsmath, epsfig}
\usepackage{changepage}
\usepackage{caption}
\usepackage[section]{placeins}
\makeatletter
\AtBeginDocument{%
  \expandafter\renewcommand\expandafter\subsection\expandafter{%
    \expandafter\@fb@secFB\subsection
  }%
}
\makeatother

\def\p{\partial}

\def\nh{\noindent\hangindent=1 true cm \hangafter = 1}

\def\EX{{\rm E}}
\def\nh{\noindent\hangindent=1 true cm \hangafter = 1}

\def\B {\begin{eqnarray*}}

\newcommand{\bel}[1]{\begin{equation}\label{#1}}

\newcommand{\be}{\begin{equation}}
\newcommand{\qe}{\end{equation}}
\newcommand{\ee}{\end{equation}}
\newcommand{\baS}{\begin{eqnarray}}
\newcommand{\ba}{\begin{eqnarray}}

\newcommand{\ea}{\end{eqnarray}}

\def\EN{\end{eqnarray*}}

\begin{document}
\title{Density independent population growth with random survival\\
\   \\
{\normalsize
Henry C. Tuckwell$^{1,2\dagger, *}$\\   \
\  \\ 
 \              \\
$^1$ School of Electrical and Electronic Engineering, University of Adelaide,\\
Adelaide, South Australia 5005, Australia \\
 \              \\
$^2$ School of Mathematical Sciences, Monash University, Clayton,
Victoria 3800, Australia \\
\  \\
$^{\dagger}$ {\it Email:} henry.tuckwell@adelaide.edu.au\\ 
\     \\}}

\maketitle

%
%
\begin{abstract}  
A simplified model for the growth of a population
is studied in which random effects arise because 
reproducing individuals have a certain probability of surviving until the
next breeding season and hence contributing to the next generation. 
The resulting Markov chain is that of a branching process with a known
generating function. For parameter values leading to non-extinction,
an approximating diffusion process is obtained for the population size. 
 Results are obtained for the number of
offspring $r_h$ and the initial population size $N_0$ required to guarantee a given probabilty of survival. For large probabilities of survival, increasing
the initial population size from $N_0=1$ to $N_0=2$ gives a very
large decrease in required fecundity but further increases in $N_0$ lead to much smaller decreases in $r_h$.  For small probabilities  (< 0.2) of survival the decreases in required fecundity when $N_0$ changes from 1 to 2  are very small.  The calculations have relevance to the survival of populations derived from colonizing individuals which could be any of a variety of organisms.

 \end{abstract}
\centerline {{\it Keywords}: Population ecology; viability; extinction
probability. }

\newpage
\rule{60mm}{1.5pt}
\tableofcontents
\rule{60mm}{1.5pt}
%
\newpage
  \section{Introduction}
The temporal and spatial dynamics of the growth or decline and spread  
of populations of animals, plants, cells and microorganisms such as bacteria and viruses involves a large number of factors which is mostly species and environment specific.  Mathematical models for population growth are thus unlikely to have universal applicability but in some cases
may elucidate general principles.  The collection of experimental data
is mostly a painstaking and sometimes costly task so comparison of data with model predictions is made in relatively few instances.  Factors affecting the probability of extinction of a population
is one of the main topics of interest in theoretical ecology or theoretical population dynamics  and is the subject of the present article.

There is an immense number of population growth models that have been developed since the pioneering models of  exponential growth Malthus (1798) and logistic growth
Verhulst (1838).  See for example Lefever and Horsthemke (1979),
Cooke and Witten (1986), Collins and Glenn (1991), Tuckwell and Le Corfec (1998), Tuckwell et al. (2008) and Ferguson and Ponciano (2015). Some of these works concern  models with spatial properties and some contain the analysis of experimental data.  See
Vandermeer and Goldberg (2013) for a useful review.

One of the first models for the random growth of populations due to 
fluctations in environmental factors was that of  Lewontin and Cohen (1969).  Although random processes had been employed and analyzed in population genetics since the 1920s (Wright, 1931) and particularly with diffusion approximations (Kimura, 1964; Crow and Kimura, 1970),   it was not
until the 1970s that the theory and analysis of stochastic differential equations, 
developed by Ito (1951) and Kolmogorov (1938),  began to be applied
in biological modeling.  Such coincided with the appearance of many
expository texts on stochastic processes and their applications (e.g., Cox and Miller, 1965; Jaswinski, 1970; Gihman and Skorohod, 1972)

One of the models for density independent population growth advanced by Levins (1969)
for a population of 
size $N(t)$ at time $t$, with initial value $N_0 \in (0,\infty)$
took the form of a differential equation
$$ \frac{dN}{dt}=r(t)N + \epsilon(t) \sqrt{v(1-v)N},$$
where $r$ is the intrinsic growth rate, $v$ is called the mean viability
and $\epsilon(t)$ is  ``a random variable with mean 0 and variance 1''.
According to the description of that model   `` the sampling variation occurs only in the death of adults''. 
A related stochastic differential equation is 
\be     dN=rNdt + \sqrt{v(1-v) N}dW \ee
where $r$ is the instrinsic growth rate, $W$ is a standard 1-parameter
Wiener process with mean zero and $Var[W(t)]=t$. This stochastic
equation was investigated fully in Tuckwell (1974) who found the
associated transition probability density function and the probability of extinction, $P_E$.
The latter was, given by
\be  P_E=1 \ee
if $r \leq 0$ and by
\be P_E= 2\mathcal{N}\bigg(-2\sqrt \frac {rN_0}{v(1-v)}  \bigg)  \ee
when $r>0$, $\mathcal{N}$ being the normal distribution function. 

\section{Description of model}

The model considered in this article has the following assumptions.
\begin{itemize}
\item{The population has a breeding season (or seasons)}.
\item{Only reproducing individuals (often females who can reproduce) are taken into account in enumerating the population size.}
\item{A reproducing individual (female) who is alive at the start of the breeding season produces
  $r$ reproducing (female) offspring.}
\item{The probability that a reproducing individual survives to the next
breeding season is $v$.}
\item{Survival of an individual to the next breeding season is independent
of the survival of any other individual.}
\end{itemize}

In this model, the growth of the population is partly deterministic (the birth process) and partly governed by random influences (survival to the next
breedong season). Note that the number of offspring can easily be made a random variable,
but for simplicity this is ignored here.  The fact that the number of offspring is the same for all reproducing individuals can be interpreted by assuming that the fecundity $r$ has been averaged across and within
age groups within the population. This averaging should not produce any significant  qualitative or quantitative effects. The same can be said to
apply to the averaging of survivorship $v$ over and within age groups. 

Note also that in some populations of wild animals, not all females, even if of reproductive age,  are allowed 
to reproduce by the dictates of senior females.  In fact, this applies to most human populations as young women are not encouraged to
reproduce until they achieve a certain age or social status. 
The $r$  (female) offspring per individual
may not become reproducing in the breeding season after their
birth so a time delay could be introduced to allow for this effect.
However, this complication is not taken into account here, with appeal
to an averaging argument. 
Furthermore, there may be an upper limit to  the age at which
females can reproduce, such being associated with menopause
in human and certain other mammalian populations.  In the
present model passing this age is equivalent to death so that $v$
is actually the probability that a reproducing female not only
survives to the next generation but also continues to contribute 
offspring, unimpeded by either social rules or by aging.

\section{Analysis of model}
Reproducing females will hereafter be referred to simply as 
individuals. 
Suppose that just prior to a certain breeding season there are $N_0$
individuals. According to the above assumptions, the expectation of the number of individuals $N_1$ just prior to the next breeding season will be
\be \EX[N_1|N_0]=v(1+r)N_0, \ee
and the variance of this quantity is
\be {\rm Var}[N_1|N_0] = v(1-v)(1+r)N_0. \ee
It is seen that the model is equivalent to a Markov chain with transition
probabilities
\be P_{N_0, N_0(r+1)-j} = {N_0(r+1) \choose {N_0(r+1)-j }}v^{N_0(r+1)-j}(1-v)^j, \ee
where $j=0,1,..., N_0(r+1)$. 
In particular, the Markov chain constitutes a branching process, which, in the standard notation of Feller (1968, p 295) has the generating function
\be P(s)= \sum_{k=0}^{1+r} {1+r \choose k}(vs)^k(1-v)^{1+r-k}, \ee
where the number of individuals in the zeroth generation is 1. 

Before applying the results for branching processes, we note that
if $v=0$ then the population must go extinct. If at the other extreme,
if $v=1$ the process is one whose continuous approximation grows according to the Malthusian law
\be \frac{dN}{dt}=rN. \ee
If we assume that in Levins' model, the parameter $v$ is in fact
a probability,  then these extreme values of $v$ do not yield any difference in the resulting growth process, because setting $v=0$ or $v=1$ in the stochastic differential equation  (1) yields equation (8).

For the branching process with generating function (7) the following exact
results hold.  Firstly, the expectation of the population size at the $n$-th
generatiion is 
\be \EX[N_n|N_0=1]= (v(1+r))^n. \ee
Secondly, the variance is 
\be   {\rm Var}[N_n|N_0=1] = (1-v)(v(1+r))^n \sum_{k=0}^{n-1} (v(1+r))^k. \ee
Finally, the population is bound to go extinct if
\be  r \leq \frac{1}{v} -1=r_c, \ee
$r_c$ denoting a crtical value. 

Furthermore, it is a well-known result for branching processes such as 
that considered here, that if the population survives then the population
size eventually becomes very large. Hence, in the biologically interesting case when condition (11) is violated, we can study the evolution of the 
population by means of the corresponding diffusion, approximation. This approach has been widely used in the study of the evolution of gene frequencies (Crow and Kimura, 1970). Thus, though it is possible to write down the equations for the probability of extinction in our model in terms of the generating function of the branching process, this quantity may be found more readily as a function of $v$, $r$ and $N_0$ through the use of a continuous Markov approximation.

\subsection{A diffusion approximation}
Let $N(t)$ be the continuous approximation to the population size at time $t$ and let $f(N,t|N_0)$ be the transition probability density defined through
\be f(N,t|N_0)dN = {\rm Pr}[N(t) \in (N, N+dN) |N(0)=N_0]. \ee
For a continuous Markov process (diffusion) the transition density
satisfies a forward Kolmogorov equation
\be \frac{\p f}{\p t} = - \frac{\p }{\p N} \big[K_1(N)f\big] + \frac{1}{2}
   \frac{\p^2}{\p N^2} \big[K_2(N)f \big],\ee
where $K_1(N)$ and $K_2(N)$ are the first and second infinitesimal
moments. According to the above reference (Crow and Kimura, 1970)
we should, for the diffusion approximation to our branching process, set
\be K_1(N)= v(1+r)N \ee
and
\be K_2(N)= v(1-v)(1+r)N \ee
where these two quantities have been obtained from equations (4) and (5). The solution of equation (13) with infinitesimal moments given by (14) and (15) has been obtained as an infinite series by Feller (1951).

Let us put
\be \alpha = v(1+r)-1.  \ee
 Then from that author's results we obtain the following expressions for
the mean and variance of $N(t)$.
\be \EX[N(t)|N(0)=N_0]=N_0  \exp [\alpha t] \ee
Secondly, the variance is 
\be   {\rm Var}[N(t)|N(0)=N_0] = \frac{N_0}{\alpha} v(1-v)(1+r)
  \exp[\alpha t] \big( \exp[\alpha t] -1\big).  \ee 
Furthermore, the probability that the population eventually goes extinct is

\begin{equation}
P_E =
\begin{cases}
1, & \text{if }  r \leq \frac{1}{v}-1,\\
\exp\big[ \frac{-2\alpha N_0}{v(1-v)(1+r)}\big], & \text{if }  r > \frac{1}{v}-1.\\
\end{cases}
\end{equation}
Hence the exact results for the branching process and those for the
diffusion approximation agree insofar as the conditions for certain extinction are the same for both.


The diffusion process represented by the Kolmogorov equation
for the present model can be considered to be associated with
the stochastic differential equation
\be dN = (v(1+r)-1)Ndt + \sqrt{v(1-v)(1+r)N}dW. \ee
This equation has the same form as (1) but with different drift
and diffusion coefficients. Note that the result of associating a Kolmogorov equation with a stochastic differential euquation 
depends on which definition of stochastic integral is employed, two
common examples being those of Ito and Stratonovich (see e.g., 
Mortensen, 1969).

If we put $v=1$ in Equ. (20), meaning that the entire population
survives from breeding season to breeding season, then we recover
the Malthusian growth equation (8), as expected. However, this
iterative process could not occur indefinitely because it assumes that
individuals may persist indefinitely. On the other hand, if we set $v=0$
we obtain the solution $N(t)=N_0\exp(-t)$, which means that the
population is destined for extinction. In all cases there is a 
contribution from both the viability $v$ and the net growth parameter $r$
which sees to be a desirable feature in a model describing  the evolution of a population whose members have a certain probability of surviving to
the next breeding season. The parameter $v$ can take into account
age-dependent death rates when it is between $0$ and $1$ so that the case  $v=1$ and concomitant Malthusian growth does not occur.

\subsection{Examples}
We may use the above results for the diffusion approximation to
study the properties of those populations which in the branching 
process model have a positive probability of ultimate survival.
We observe that in the branching process model the number of offspring per individual is restricted to positive integer values.

In populations not bound for certain extincton it is apparent from (17) and (18) that the expectation and variance of $N(t)$  become infinite as $t \rightarrow \infty$, which is also true for the branching process.
It is also clear that there is no value of $r \in (0, \infty)$ which leads to certain survival if $0 \leq v < 1$. That is, unless the probability of an individual's survival to the next breeding season is unity, then no matter
how great the number of offspring per individual per breeding season, 
there is always a non-zero chance of ultimate extinction of the population.

Our main interest here is to ascertain quantitative estimates of the effects of the parameters $v$, the viability, and $r$, the fecundity,
on survival. These estimates will mainly be presented graphically.
Recall that  $r_c$ is the critical value of $r$ defined in (11). We also define the number of offspring $r_h(v, P_E)$ which gives a 
probability $P_E$ of ultimate extinction when the viability is $v$.  
This is given by the expression
\be
r_h(v,P_E)=   \frac {v^{-1}}{  1 +  \frac{(1-v) \ln(P_E) } {2N_0 }     } -1. \ee
Equivalently this gives a probabilty of ultimate survival
\be P_S=1-P_E. \ee
Some algebra shows that apart from a singularity at $v=0$,
 $r_h$ has another singularity at 
\be v^*=1 + \frac{2N_0}{\ln P_E}. \ee
The second singularity only occurs at positive values if 
\be P_E < \exp[-2N_0] \doteq P_E^*, \ee
so formula (21) is only valid when $P_E > P_E^*$. 
The smallest initial population is technically $N_0=1$ for which the requirement for a non singular $r_h$ is $P_E>\exp[-2] \approx 0.1353$.
However, for $N_0=2$ the value of $P_E^* \approx 0.0183$ and for $N_0=3$ it is $P_E^* \approx 0.0025$, so that only in very few
cases is there a singularity for positive values of $v$.

Similarly, another rearrangement of (19) gives the value of the viability $v_h(r, P_E)$
which gives a probability  $P_E$ of extinction when the 
fecundity is $r$, 
\be v_h(r,P_E)=\frac{1}{2\gamma}\bigg(\gamma-2N_0+ \sqrt\bigg[
(\gamma-2N_0)^2 + \frac{8\gamma N_0}{1+r} \bigg] \bigg), \ee
where $\gamma=-\log P_E$.

\subsubsection{Graphical results}
Figures 1 and 2 show plots of $r_h$ versus $v$ for four values of the probability of ultimate survival $P_S=0.1, 0.15, 0.5, 0.85$ and for four values of the initial number of reproducing females $N_0=1,2, 10, 100$.  Values of $P_S$ are indicated on each plot, but ony the two extreme values of $N_0$ are indicated on the plot for $P_S=0.85$. In order to make the pictures clearer, in Figure 1 values of $v$ are small, being less than 0.15, whereas in Figure 2 they are large with $0.15 \leq v \leq 1$. In all calculations of $r_h$, only values of $v$ greater than or equal to 0.01 were considered in order to stay away from the singularity at $v=0$. In Figure 1 are also shown the critical values $r_c$ (diamonds) from (11) below which value of $r$ extinction is certain. 
This curve is almost coincident with the values of $r_h$ for very large
initial population sizes as 
\be \lim_{N_0 \rightarrow \infty}
 r_h = r_c. 
\ee

The most noticeable features of the results shown in Figures 1 and 2 are as follows.

\begin{itemize}
\item{When the probability of survival is large, increasing the initial
population size from $N_0=1$ to $N_0=2$ results, for a given
$v$,  in a very large decrease in the required fecundity $r_h$. For example, with $P_S=0.85$ and $v=0.1$, increasing $N_0$ from
1 to 2 gives rise to a drop in $r_h$ from about 65 to less than 20.
Further increases in $N_0$ do not lead to very large changes in
$r_h$.}
\item{The smaller the probability of survival, at a given value of $v$,  the smaller the 
change in required $r_h$ when $N_0$ increases from 1 to 2.}
\item{When the probability of survival is small, for a given $v$, changing the initial population size has very little effect on the required fecundity $r_h$ - see for example the results for $P_S=0.10$ in Figure 1. }
\item{Regardless of the probability of survival or the initial population
size,  $r_h$ decreases as $v$ increases in an exponential-like fashion.}
\end{itemize}

     \begin{figure}[!h]
\begin{center}
\centerline\leavevmode\epsfig{file=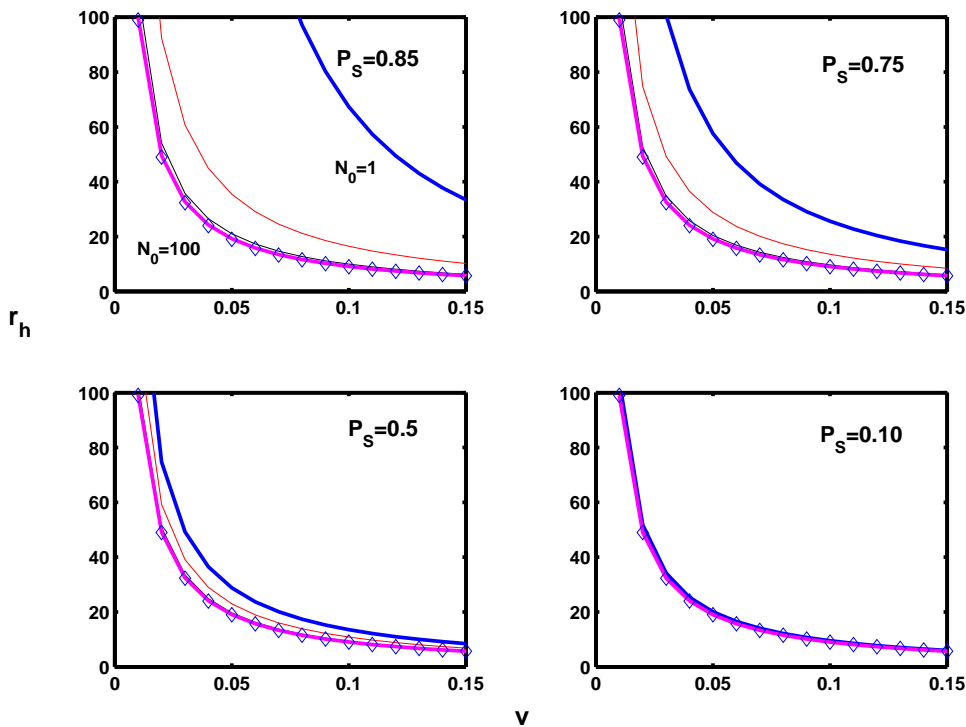,width=5in}
\end{center}
\caption{Showing the number $r_h$ of offspring per individual per breeding season as a function of the survival probability $v$ of individuals
for four values of the probability $P_S$ of ultimate survival and for four values of the initial population size $N_0$. In this figure results are
restricted to small values of $v \in [0.01, 0.15]$.  The diamonds indicate
the values of $r_c$ from (11) below which extinction is certain. 
 } 
\label{fig:wedge}
\end{figure}

     \begin{figure}[!h]
\begin{center}
\centerline\leavevmode\epsfig{file=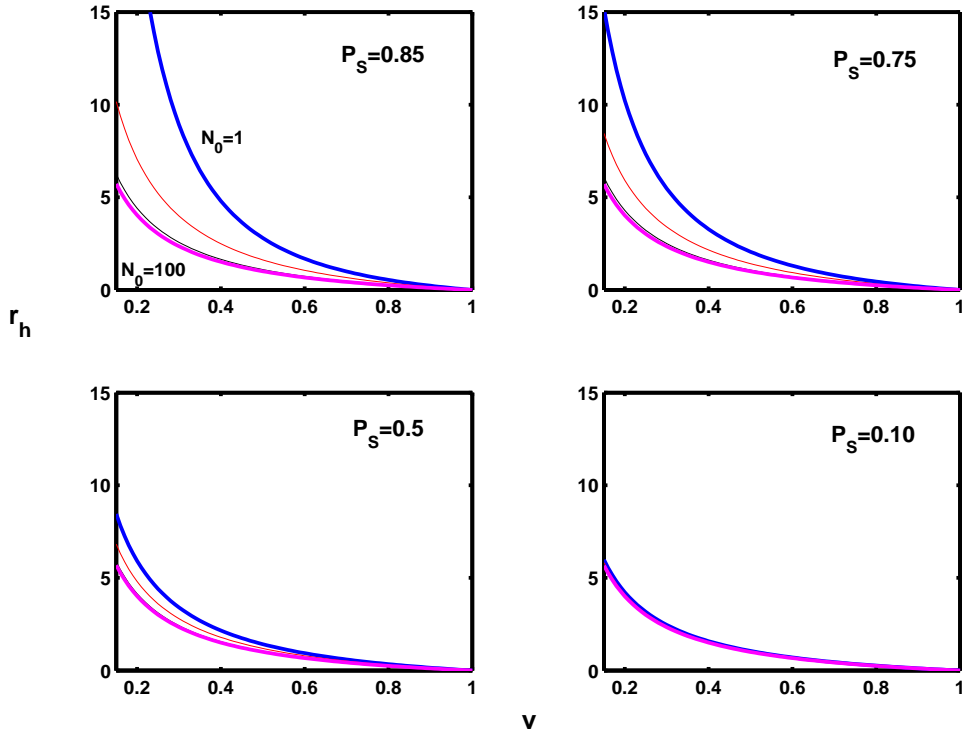,width=5in}
\end{center}
\caption{As in the previous figure but that $v \in [0.15, 1.0]$.
 } 
\label{fig:wedge}
\end{figure}

It is of interest to examine the dependence of the probability of extinction,
$P_E$, on the fecundity for various values of the survival probability $v$
and the initial population size $N_0$. Figure 3 shows $P_E$ versus 
$r$ for three values of $v$=0.05, 0.15 and 0.5 (blue, red and black curves respectively)  and for three values of $N_0$=1,2 and 3 (solid, dashed and dot-dash curves respectively).  

\begin{itemize}
\item{For the population to have any chance of survival $r$ must be greater than the critical value
$r_c$ given by (11).}
\item{ When $v$ is small $P_E$ is unity until $r$ is large
as in the case $v=0.05$ where $P_E=1$ until $r=19$. The rate of decline in $P_E$ for larger $r$ is slowest  with an initial population of $N_0=1$,
becoming faster as $N_0$ increases. By $r=30$, the values of $P_E$
have fallen to approximately 0.49, 0.22 and 0.11  for initial populations
of 1,2 and 3 respectively.}
\item{When $v$ is large, $P_E$ drops dramatically after $r$ increases beyond the critical value. For example, when $v=0.5$ the critical value of $r_c=1$. With $r=3$ the values of $P_E$ are about 0.14, 0.02 and 0.003 for $N_0$=1,2 and 3 respectively, representing jumps from zero
probability of ultimate survival to 86\%, 98\% and 99.7\% by an increase
from 1 to three offspring per reproducing female. }
\end{itemize}

     \begin{figure}[!h]
\begin{center}
\centerline\leavevmode\epsfig{file=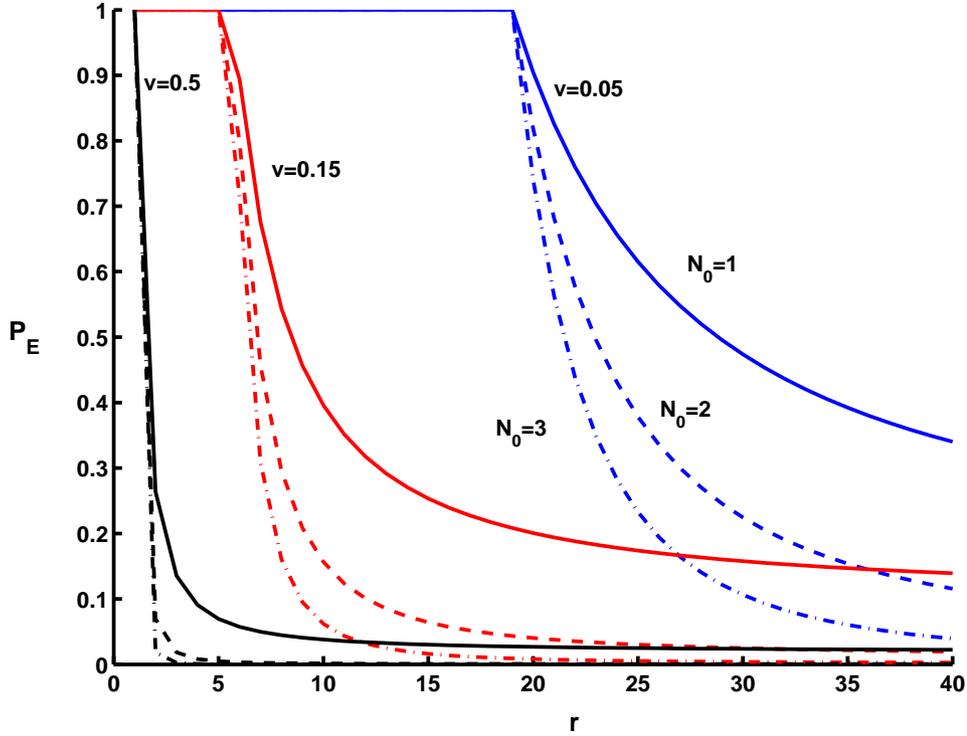,width=5in}
\end{center}
\caption{Plots of probability of ultimate extinction versus $r$ for various values of survival probabilty $v$ and initial population size $N_0$. Blue, red and black curves for values of $v$ which are small (0.05), intermediate (0.15) and large (0.5), respectively.  Solid, dashed and dot-dashed curves for initial populations of 1, 2 and 3 respectively. 
 } 
\label{fig:wedge}
\end{figure}

Table 1 contains calculated values for $P_E$ for a wide range
of values of $v$ with two values of $r$ for each and with 4 values
of $N_0$. Only when $v$ is small 	(0.01 or 0.05) is $P_E$ substantial for any of the given values of $r$ and $N_0$.   When $v=0.95$ the probabilities of extinction are always essentially zero, even when $N_0=1$.

\begin{center}

\begin{table}[h]
    \caption{Numerical values for $P_E$=extinction probability}
\smallskip
\begin{center}
\begin{tabular}{ccccccc}
\hline 
$v$   & $r_c$  & $r$ & $N_0=1$  & $N_0=10$ & $N_0=100$ & $N_0=1000$ \\
  \hline
0.01 & 99 & 100 & 0.9802 & 0.8187  &  0.1353 & 2.1x10-9 \\
0.01 & 99 & 200 & 0.3624 & 3.9x10-5  & 8.2x10-45 & $\approx$0 \\
0.05 & 19 & 20 & 0.9046 & 0.3670  & 4.3x10-5 & 2.9x10-44\\
0.05 & 19 & 50 & 0.2781 & 2.8x10-6 & 2.7x10-56 & $\approx$0 \\
0.95 & 0.053 & 1 & 5.9x10-9 & 5.2x10-83 & $\approx$0  & $\approx$0 \\ 
0.95 & 0.053 & 10 & 2.0x10-16 & 8.1x10-158 & $\approx$0& $\approx$0 \\ 
 \hline
\end{tabular}

\end{center}

\end{table}

\end{center}

\section{Summary and conclusions}
We have considered a simple approximate model for population growth
in which reproducing females, which must be carefully defined,
produce $r$ offspring, and 
have a probability $v$ of surviving to the next breeding season.
Using results from branching process theory and that of diffusion processes,
the probability of survival is obtained in terms of the parameters
$r$ and $v$ and the initial population size. Numerical results are
presented both graphically and in tablular form. When the probability of ultimate survival of the population is large, small increases in $N_0$ change the fecundity required by large amounts whereas when 
the probability of survival is small, the initial population size is found to have little influence. These calculations have relevance to the survival of colonizing individuals which could be from populations of animals, plants, 
insects, cells or microorganisms such as bacteria and viruses. 

\section{References}
\nh Collins, S.L., Glenn, S.M., 1991.  Importance of spatial and
temporal dynamics in species regional abundance and distribution.
Ecology 72, 654-664.

\nh Cooke, K.L., Witten, M., 1986. 
One-dimensional linear and logistic
harvesting models. Math. Modeling 7, 301-340.

\nh Crow, J.F., Kimura, M., 1970. An introduction to population genetics theory. Harper and Row, New York. 

\nh Feller, W., 1951. Diffusion processes in genetics. Proc. 2nd Berkeley Symp. Math. Stat. Prob., 227-246. Univ. California Press,
Berkeley.

\nh Feller, W., 1968. An introduction to probability theory and its applications. Wiley, New York. 

\nh Ferguson, J.M., Ponciano, J.M., 2015. 
Evidence and implications of higher-order scaling
in the environmental variation of animal
population growth. PNAS 112, 2782-2787. 

\nh Gihman, I.I., Skorohod, A.V., 1972. Stochastic differential equations. Springer, Berlin.

\nh Ito, K., 1951. On stochastic differential equations. Mem. Amer. Math. Soc. 4.  

\nh Jaswinski, A.H., 1970. Stochastic processes and filtering theory.
Academic Press, New York.

\nh Kimura, M., 1964. Diffusion models in population genetics.
  J. Appl. Prob. 1, 177-232. 

\nh  Kolmogorov, A.N., 1938. On the analytic methods of probability theory. Uspekhi matematicheskikh nauk 5, 5-41.

\nh Lefever, R., Horsthemke,W., 1979.  Bistability in fluctuating environments.
Implications in tumor immunology. Bull. Math. Biol. 41, 469-490.

\nh Levins, R., 1969.  The effect of random variations of different
types on population growth. PNAS 62, 1061-1065.

\nh Lewontin, R.C., Cohen, D., 1969.
On population growth in a randomly varying
environment.  PNAS 62, 1056-1060.

\nh Malthus, T.R., 1798. An Essay on the Principle of Population, as it affects the Future Improvement of Society: with remarks on the Speculations of Mr. Godwin, M. Condorset, and other writers. J. Johnson, London.

\nh Mortensen, R.E., 1969. Mathematical problems of modeling
stochastic nonlinear systems. J. Stat. Phys. 1, 271-296. 

 \nh Tuckwell, H.C., 1974. A study of some diffusion models of
population growth. Theor. Pop. Biol. 5, 345-357. 

\nh Tuckwell, H.C., Le Corfec, E., 1998. A stochastic model for early HIV-1 population dynamics. J Theor. Biol. 195, 451-463. 

\nh Tuckwell, H.C., Shipman, P.D., Perelson, A.S., 2008. 
The probability of HIV infection in a new host and its reduction with
microbicides. Math. Biosci. 214, 81-86. 

\nh  Vandermeer, J. H., Goldberg, D.E.,  2013.  Population Ecology:  First Principles.  Princeton Univ. Press, Princeton, NJ.

\nh Verhulst, P. F., 1838. Notice sur la Loi que la Population suit dans son Accroissement. Cor mathematique et physique 10, 113-121.

\nh  Wright, S., 1931. Evolution in Mendelian populations. Genetics 16, pp.97-159.

\end{document}